\documentclass[twocolumn,preprintnumbers,superscriptaddress
,footinbib
]{revtex4-1}

\usepackage{simplewick}

\newcommand{\beq}{\begin{eqnarray}}
\newcommand{\eeq}{\end{eqnarray}}

\usepackage{graphicx}
\usepackage{amssymb}
\usepackage{amsmath}
\usepackage{color}
\usepackage{epsfig}
\usepackage{epstopdf}
\usepackage{hyperref}
\usepackage{subfigure}

\allowdisplaybreaks

\usepackage{mathtools}

\DeclarePairedDelimiterX\braket[2]{\langle}{\rangle}{#1 \delimsize\vert #2}

\begin{document}

\preprint{RIKEN-iTHEMS-Report-22}
\date{\today}
\title{Velocity of Sound beyond the High-Density Relativistic Limit from Lattice Simulation of Dense Two-Color QCD
}

\author{Kei Iida}
\email[]{iida(at)kochi-u.ac.j}
\affiliation{{\it
Department of Mathematics and Physics, Kochi University, 2-5-1 Akebono-cho, Kochi 780-8520, Japan
}}

\author{Etsuko Itou}
\email[]{itou(at)yukawa.kyoto-u.ac.jp}
\affiliation{\it {Interdisciplinary Theoretical and Mathematical Sciences Program (iTHEMS), RIKEN, Wako 351-0198, Japan}}
\affiliation{\it {Department of Physics, and Research and 
Education Center for Natural Sciences, Keio University, 4-1-1 Hiyoshi, Yokohama, Kanagawa 223-8521, Japan}}
\affiliation{\it {Research Center for Nuclear Physics (RCNP), Osaka University, Osaka 567-0047, Japan}}

\begin{abstract}

We obtain the equation of state (EoS) for two-color QCD at low temperature and high density  from the lattice Monte Carlo simulation. We find that  the velocity of sound exceeds the relativistic limit ($c_s^2/c^2=1/3$) after BEC-BCS crossover in the superfluid phase.
Such an excess of the sound velocity is previously unknown from any lattice calculations for QCD-like theories.
This finding might have a possible relevance to the EoS of neutron star matter revealed by recent measurements of neutron star masses and radii.

\end{abstract}
\maketitle

\renewcommand{\thefootnote}{\arabic{footnote}}
\setcounter{footnote}{0}

\newpage


\setcounter{page}{1}

The equation of state (EoS) of dense QCD at low temperature is  still poorly known but is indispensable
 particularly because it is related with understanding  neutron star observations including recent simultaneous measurements of masses and radii of neutron stars ~\cite{Alford:2007xm,Masuda2013-jk,Watts2016-be,Baym:2017whm,LIGOScientific:2018cki,Huth:2021bsp,Kojo2021-on}.
Several early works based on 
 a phenomenological quark-hadron crossover picture of neutron star matter~\cite{Masuda2013-jk,Baym:2017whm} suggested that
 the  zero-temperature sound velocity squared, $c_s^2=\partial p/\partial e$, peaks in the intermediate density region  in such a way as to fulfill various observational constraints.
Here, $p$ and $e$ denote the pressure and internal energy density of the system, respectively.
More recently, based on a quarkyonic matter model, McLerran and Reddy~\cite{McLerran2019-qh} have shown that the peak appears slightly above nuclear saturation density.
Furthermore, Kojo ~\cite{Kojo2021-mg} proposed a microscopic interpretation on the origin of the peak 
based on a quark saturation mechanism, which is supposed to work for any number of colors. Actually, Kojo and Suenaga~\cite{Kojo2021-wh} indicated that a similar peak  of $c_s^2$ emerges not only in $3$-color QCD, but also in $2$-color QCD.

The intermediate density regime, which intervenes
between the dilute hadron and perturbative QCD (pQCD) regimes,  is not analytically accessible.
The first-principles calculations of dense QCD have been desired,
but not yet been successful because of the severe sign problem.
On the other hand, the sign problem is absent in even-flavor dense $2$-color QCD  because of the pseudo-reality of fundamental quarks.
In the case of $2$-color QCD,  furthermore, the diquark condensate, which occurs in the superfluid phase, is color singlet. 
Then we can add  an external source term of the diquark condensate to explicitly break the U(1) baryon symmetry as a standard technique to study spontaneous symmetry breaking.
It allows us to perform numerical simulations of $2$-color QCD in the superfluid phase without any approximation.
$2$-color QCD at zero chemical potential exhibits the same properties  as $3$-color QCD, {\it e.g.,} confinement, spontaneous chiral symmetry breaking, and 
thermodynamic behaviors.
Under these circumstances, it is expected that $2$-color QCD even at non-zero chemical potential  could be a good testing ground  in obtaining qualitative understanding in dense QCD.

Based on this motivation, several Monte Carlo studies on $2$-color QCD have been conducted  independently and intensively in recent years~\cite{ Muroya2002-eh,Muroya2002-qc,Hands2006-mh,Hands2007-vp,  Hands2011-jh, Cotter2012-bh,  Hands2012-fn,Cotter2012-zl,Boz2013-pz, Boz2015-ex,Braguta2016-ds, Itou2018-py, Astrakhantsev:2018uzd, Boz2019-fl,Boz2019-uz,Iida:2019rah,  Astrakhantsev2020-wi,Buividovich2020-ld,  Iida:2020emi,Ishiguro:2021yxr,Bornyakov2022-sv}.
Putting the results from Refs.~\cite{Cotter2012-bh, Cotter2012-zl, Braguta2016-ds, Iida:2019rah, Boz2019-fl, Bornyakov2022-sv} together, one can conclude that the $2$-color QCD phase diagram is quantitatively  clarified.
Most remarkably, the emergence of superfluidity at fairly high temperature, $T\approx 100$ MeV, has been found.

In this work, we numerically obtain the EoS and the sound velocity in dense $2$-color QCD.  
 We use the same lattice setup as our previous works~\cite{Itou2018-py,Iida:2019rah,Iida:2020emi,Ishiguro:2021yxr} and confine ourselves to $T\approx 79$ MeV, where the hadronic, hadronic-matter, BEC (Bose-Einstein condensed), and BCS phases emerge as density increases.
Although  first-principles calculations of EoS
have been performed in~\cite{Hands2006-mh,Hands2012-fn,Boz2019-uz,Bornyakov2022-sv}, the sound velocity has not yet been examined.

Let us explain our simulation strategy.
The lattice gauge action used in this work is the Iwasaki gauge action, which is composed of the plaquette term with $W^{1\times 1}_{\mu\nu}$ and the rectangular term with $W^{1\times 2}_{\mu\nu}$,  
\beq
S_g = \beta \sum_x \left(
 c_0 \sum^{4}_{\substack{\mu<\nu \\ \mu,\nu=1}} W^{1\times 1}_{\mu\nu}(x) +
 c_1 \sum^{4}_{\substack{\mu\neq\nu \\ \mu,\nu=1}} W^{1\times 2}_{\mu\nu}(x) \right) ,\nonumber\\
 \label{eq:gauge-action}
\eeq
where $\beta=4/g_0^2$ in the $2$-color theory and $g_0$ denotes the bare gauge coupling constant.
Under the normalization condition $c_0+8c_1=1$, the coefficient $c_1$ is set to $-0.331$~\cite{Iwasaki:1983iya}.

The two-flavor fermion action including the quark number operator and the diquark source term is given by
\beq
S_F&=& (\bar{\psi_{1}} ~~ \bar{\varphi}) \left( 
\begin{array}{cc}
\Delta(\mu) & J \gamma_5 \\
-J \gamma_5 & \Delta(-\mu) 
\end{array}
\right)
\left( 
\begin{array}{c}
\psi_{1}  \\
\varphi  
\end{array}
\right)
 \equiv  \bar{\Psi} {\mathcal M} \Psi, \nonumber\\ \label{eq:def-M}
\eeq
where
$\bar{\varphi}=-\psi_2^T C \tau_2, ~~~ \varphi=C^{-1} \tau_2 \bar{\psi}_2^T.$
Here, the indices $1,2$ of $\psi$ denote the label of the flavor and the $\Delta(\mu)_{x,y}$  is the Wilson-Dirac operator with the number operator.
The additional parameter $J$ corresponds to the diquark source parameter,
which allows us to perform the numerical simulation in the superfluid phase.
Note that $J=j \kappa$, where $j$ is a source parameter in the corresponding continuum theory, and $\kappa$ is the hopping parameter.
The $C$ in $\bar{\varphi},\varphi$ is the charge conjugation operator, and $\tau_2$ acts on the color index.
The square of the extended matrix ($\mathcal M$) can be diagonal, 
but $\det[{\mathcal M}^\dag {\mathcal M}]$ corresponds to the fermion action for the four-flavor theory, since a single $\mathcal{M}$ in Eq.\ (\ref{eq:def-M})  represents the fermion kernel of the two-flavor theory.
To reduce the number of fermions, we take the root of the extended matrix in the action.
In practice, utilizing the Rational Hybrid Monte Carlo (RHMC) algorithm, we can generate gauge configurations.

Now, we utilize a fixed scale method to obtain the EoS  at finite density~\cite{Hands2006-mh}.
The trace anomaly can be described by the beta-function of parameters and the trace part of the energy-momentum tensor. In our lattice setup, which is explicitly given by
\beq
e-3p &=& \frac{1}{N_s^3 N_\tau} \left( a \frac{d \beta}{da} |_{\mathrm{LCP}} \langle \frac{\partial S}{\partial \beta}\rangle_{sub.}  \right. \nonumber\\
&&+ a \frac{d \kappa}{da} |_{\mathrm{LCP}} \langle \frac{\partial S}{\partial \kappa} \rangle_{sub.} 
 \left. + a\frac{\partial j}{\partial a}|_{\mathrm{LCP}} \langle \frac{\partial S}{\partial j} \rangle_{sub.} \right).\nonumber\\ \label{eq:trace-anomaly}
\eeq
Here,  $a$ is the lattice spacing, and the beta-function for each parameter is evaluated at $\mu=T=0$ along the line of constant physics (LCP). Note that there is no renormalization for the quark number density as it is a conversed quantity.
We take all physical observables in the $j \rightarrow 0$ limit,  which implies that the third term in  the right side can be eliminated. 
$\langle \mathcal{O} \rangle_{sub.} (\mu) $ denotes the subtraction of the vacuum quantity. In this paper, we take $\langle \mathcal{O} \rangle_{sub.} (\mu) = \langle \mathcal{O} (\mu)  \rangle - \langle \mathcal{O} (\mu=0) \rangle $ at a fixed temperature.

In this work, we perform the simulation with $(\beta, \kappa,N_s,N_\tau)=(0.80,0.159,16,16)$.
Thanks to the scale setting function (Eq.\ (23))  and a set of $(\beta,\kappa)$ with a fixed mass ratio of pseudoscalar and vector mesons $m_{PS}/m_V$ (Table~$1$)  in Ref.~\cite{Iida:2020emi}, 
 the coefficients can be nonperturbatively determined as $a d\beta /da|_{\beta=0.80,\kappa=0.159}=-0.352$ and $a d\kappa/da |_{\beta=0.80,\kappa=0.159}=0.0282$.

The pressure can be expressed by the integral of the number density in the thermodynamic limit, namely, $p(\mu) = \int_{\mu_o}^\mu n_q (\mu') d\mu'$.
On the lattice, we perform the numerical integration to obtain $p(\mu)$. To reduce the discretization effects, we utilize the following definition proposed in Eq.(29) in Ref.~\cite{Hands2006-mh}:
\beq
\frac{p}{p_{SB}}(\mu) = \frac{\int_{\mu_o}^{\mu} d\mu' \frac{n_{SB}^{cont.}}{n_{q}^{tree}}   n^{latt.}_q(\mu')  }{\int_{\mu_o}^{\mu} d\mu' n_{SB}^{cont.} (\mu')},\label{eq:p-scheme2}
\eeq
in which the lattice numerical data ($n_q^{latt.} (\mu)$) are normalized by the quark number density on the same lattice spacing $n_q^{tree}$ which is analytically calculated by the free field propagator on the finite lattice (See Eq.~(26) in Ref.~\cite{Hands2006-mh}).  Here, we numerically calculate $n_q^{latt.}\equiv a^3 n_q= \sum_{i} \kappa \langle \bar{\psi}_i (x) (\gamma_0 -\mathbb{I}_4) e^\mu U_{4} (x) \psi_i (x+\hat{4})  + \bar{\psi}_i (x) (\gamma_0 + \mathbb{I}_4) e^{-\mu}U_4^\dag (x-\hat{4} )\psi_i (x-\hat{4})\rangle$. Furthermore, to reduce the discretization effect of the numerical integration, we take the ratio between $p(\mu)$ and its value at the Stefan-Boltzman (SB) limit ($p_{SB}(\mu)$) which is also obtained by the numerical integration of the number density of quarks in the relativistic limit, namely $n_{SB}^{cont.}=N_fN_c(\mu T^2 + \mu^3/\pi^2)/3$, where $N_f$ ($N_c$) is the number of flavors (colors).
In \eqref{eq:p-scheme2}, $\mu_o$ represents the onset scale, namely, the starting point at which $\langle n_q \rangle$ becomes nonzero as $\mu$ increases.
In the continuum theory, the pressure scales as $p_{SB}(\mu) = \int^\mu n_{SB}^{cont.}(\mu')d\mu' \approx N_fN_c \mu^4 /(12\pi^2)$ in  the high $\mu$ regime.

To study the EoS and the sound velocity, we have increased the number of  the values of $a\mu$ at intervals of $a\Delta \mu=0.05$ and also accumulated statistics ($100$--$300$ configurations) since the previous paper~\cite{Iida:2019rah}.
The statistical errors are estimated by the jackknife analysis.
According to Ref.~\cite{Iida:2020emi}, once we introduce the physical scale as $T_c=200$ MeV, where $T_c$ denotes the pseudo-critical temperature of chiral phase transition at $\mu=0$, then  our parameter set, $\beta=0.80$ and $N_\tau=16$ ($T=0.39T_c$), corresponds to $a\approx 0.17$ fm and $T\approx 79$ MeV.
The mass of lightest pseudo-scalar (PS) meson at $\mu=0$, $m_{PS}$, is still heavy in our simulations, $am_{PS}=0.6229(34)$ ($m_{PS}\approx 750 $ MeV).

\begin{figure}[htbp]
    \begin{tabular}{c}
        \includegraphics[keepaspectratio, scale=0.25]{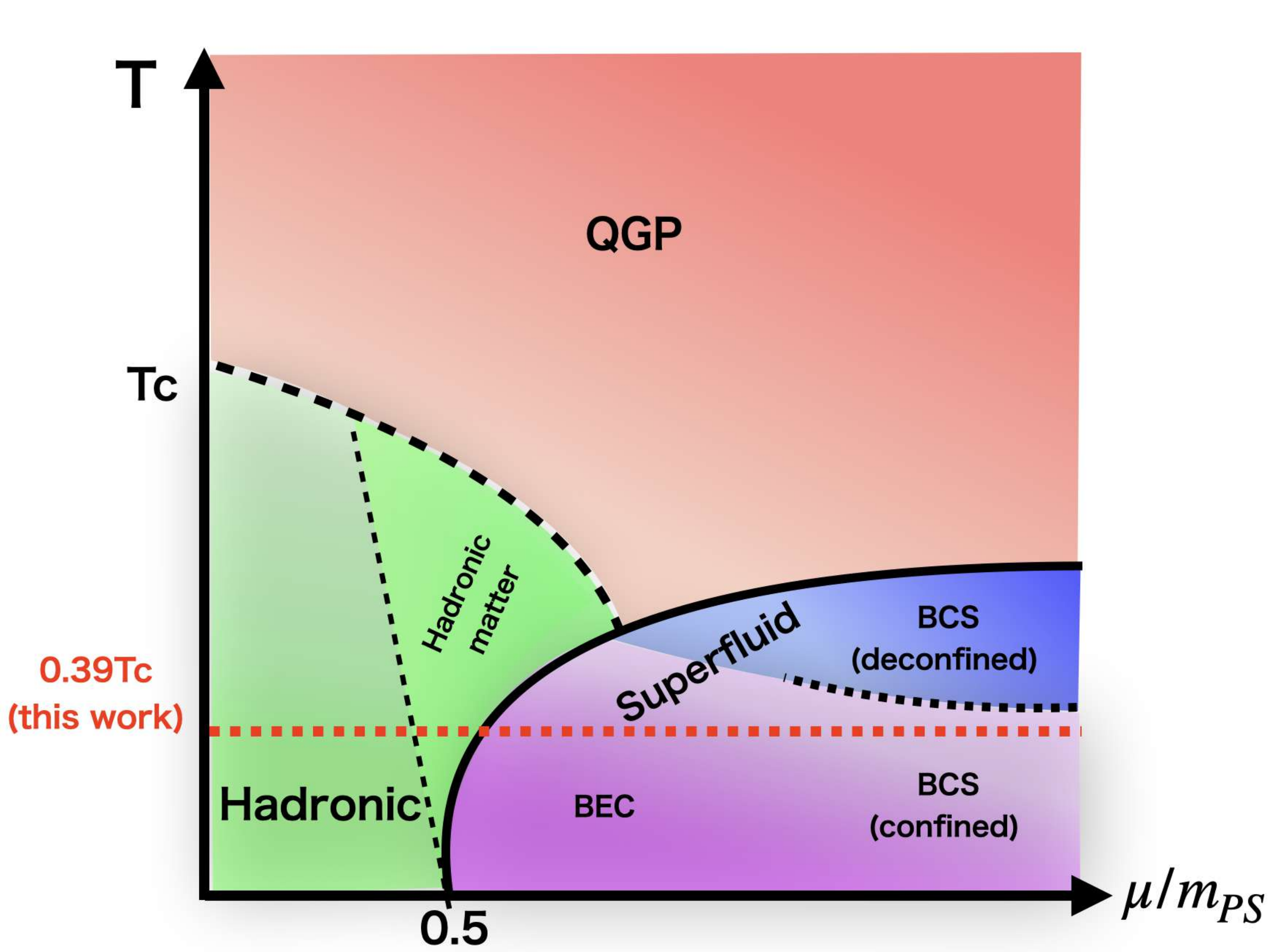}
    \end{tabular}
            \caption{
Schematic 2-color QCD phase diagram.  Each phase is defined in Table \ref{table:phase}.
}\label{fig:phase-diagram}
  \end{figure}

We show the schematic picture of the phase structure in Fig.~\ref{fig:phase-diagram} and summarize the definition of each phase in Table~\ref{table:phase}, which is  an extract from Ref.~\cite{Iida:2019rah}. 
\begin{table}[h]
\begin{center}
\begin{tabular}{|c||c|c|c|c|}
\hline
 \multicolumn{1}{|c||}{}  & \multicolumn{2}{c|}{Hadronic} &     \multicolumn{2}{c|}{Superfluid}  \\  
\cline{3-3} \cline{4-5}  & & Hadronic matter &  BEC & BCS \\  
 \hline \hline
$\langle |L| \rangle$ & zero  & zero  &    &   \\
$\langle qq \rangle$ & zero  &  zero  & non-zero & non-zero  \\ 
$ \langle n_q \rangle $ &zero &  non-zero & $  0 < \frac{\langle n^{latt.}_q \rangle}{n_q^{\mbox{tree}}} <1 $  & $  \frac{\langle n^{latt.}_q \rangle}{n_q^{\mbox{tree}}} \approx 1 $ \\ 
 \hline
\end{tabular}
\caption{ Definition of phases. } \label{table:phase}
\end{center}
\end{table}
The order parameters  that help classify the  phases are the Polyakov loop $\langle |L| \rangle$ and diquark condensate $\langle qq \rangle$, whose zero/nonzero values indicate the confinement and the superfluidity, respectively.
We found that the superfluidity emerges at $\mu_c/m_{PS} \approx 0.5$ as expected by the chiral perturbation theory (ChPT)~\cite{Kogut2000-so}.
It is natural to use $\mu/m_{PS}$ as a dimensionless parameter of density since the critical value $\mu_c$ can be approximated by $m_{PS}/2$  even if  the value of $m_{PS}$ in numerical simulation would be changed~\footnote{It is expected that the corresponding critical value of $\mu$ would be $\mu_c/m_N \approx 1/3$ if the hadronic-superfluid phase transition occurs also in the case of $3$-color QCD, where $m_N$ denotes the nucleon mass. }.
We also confirmed that the scaling law of the order parameter around it is consistent with the ChPT prediction.
Furthermore, we measured the quark number operator $\langle n^{latt.}_q\rangle$.
We identified the regime where $\langle n^{latt.}_q \rangle$ is consistent with the free quark theory as the BCS phase (See Fig.7 in Ref.~\cite{Iida:2019rah}). 
Thus, we concluded that there are hadronic, hadronic-matter, BEC and BCS phases at $T=79$ MeV, although there is no clear boundary between the BEC and BCS phases.  Interestingly,
up to $\mu/m_{PS} =1.28 $ ($\mu \lesssim 960$ MeV), the confining behavior remains,  while nontrivial instanton configurations have been discovered from calculations of the topological susceptibility~\cite{Iida:2019rah}.
It indicates that  a naive perturbative picture, for instance, pQCD, is not yet valid in the density regime studied here.

\begin{figure}[htbp]
    \includegraphics[keepaspectratio, scale=0.75]{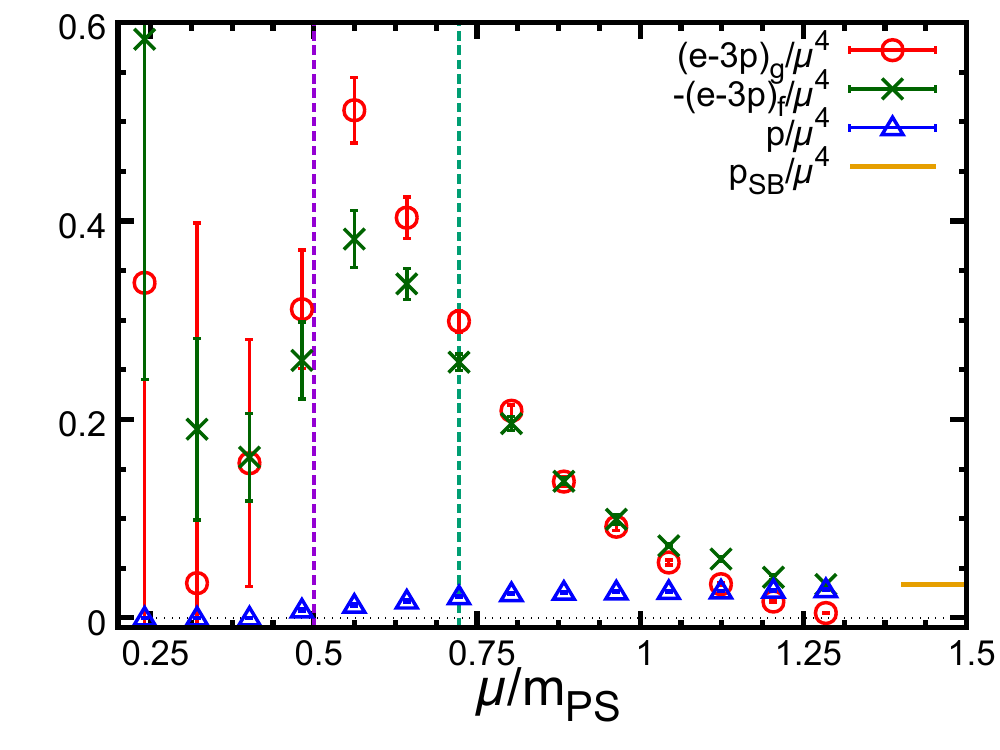}
    \caption{Trace anomaly and pressure  as a function of $\mu/m_{PS}$. The circle and cross symbols denote the gauge part and {\it minus} the fermion part of the trace anomaly, respectively. We also show $p/\mu^4$ at the relativistic limit, $p_{SB}/\mu^4=N_fN_c /(12\pi^2)$.
    The purple dashed line denotes the critical value, $\mu_c$, which is the hadronic-superfluid phase transition point, while the green dashed line indicates that the BEC-BCS crossover occurs around this value of $\mu$. }\label{fig:raw-data}
  \end{figure}

The trace anomaly and pressure are shown in Fig.~\ref{fig:raw-data}.
For the trace anomaly, we plot the gauge part (the first term in Eq.~\eqref{eq:trace-anomaly}) and minus the fermion part (the second term) separately.
Both parts are
normalized by $\mu^4$  
to see the dimensionless asymptotic  behavior.
The magnitude of each part has
a peak around the hadronic-superfluid phase transition. 
 It is very similar to the emergence of the peak of $(e -3p)/T^4$ around the hadronic-QGP phase transition at $\mu=0$.

As for the pressure, at $\mu_c=m_{PS}/2$ for the hadronic-superfluid phase transition  (purple vertical line), $p$ takes  a non-zero value since  $\langle n_q \rangle$ becomes non-zero in the hadronic-matter phase.
Thus, $\langle n_q \rangle$ becomes non-zero before the hadronic-superfluid phase transition, then  $\mu_c$ is not the same as $\mu_o$ in Eq.~\eqref{eq:p-scheme2}. 
 The low but finite temperature effects cause the discrepancy between them as discussed in~\cite{Iida:2019rah}. 
We can see that our data 
monotonically increase and approach the value in the relativistic limit.
The value of $p/p_{SB}$ is $\approx 0.84$ at the highest density in our simulation.

\begin{figure}[htbp]
  \begin{center}
    \includegraphics[keepaspectratio, scale=0.38]{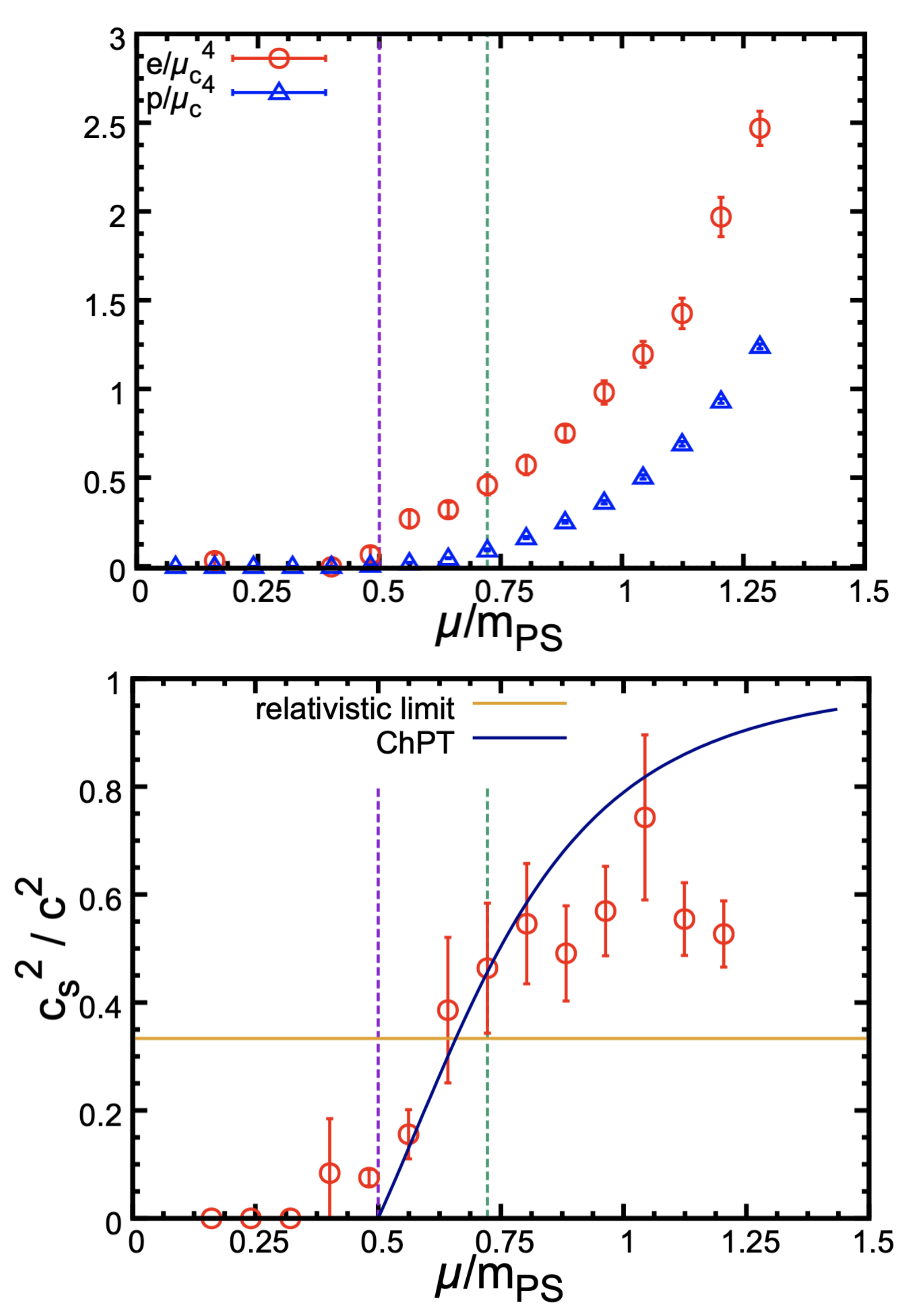}
           \caption{Top: The EoS as a function of $\mu/m_{PS}$.  Bottom: Sound velocity  squared as a function of $\mu/m_{PS}$. The horizontal line (orange) denotes the value in the relativistic limit, $c_s^2/c^2 =1/3$. The blue curve shows the result of ChPT.}\label{fig:EoS}
    \end{center}
  \end{figure}

 Combining the data of $e-3p$ and $p$ above, we finally obtain the EoS and sound velocity  as shown in the top and bottom panels of Fig.~\ref{fig:EoS}, respectively.
In the top panel, we normalize $e$ and $p$ by 
$\mu_c$ so as to be dimensionless. 
We can see that both $e$ and $p$ are consistent with zero in the hadronic phase.
Thus, these thermodynamic quantities are not changed even if $\mu$ increases before the hadronic-superfluid phase transition.

Now, let us focus on the sound velocity  depicted in the bottom panel in Fig.~\ref{fig:EoS}.
Here, we evaluate $c_s^2 (\mu)= \Delta p (\mu)/\Delta e (\mu) $, where $\Delta p (\mu)$ and $ \Delta e (\mu)$ are estimated by the symmetric finite difference, 
i.e., $\Delta p(\mu) =( p(\mu +\Delta \mu) - p(\mu -\Delta \mu))/2$.
First of all, our results are consistent with the prediction of ChPT~\cite{Son_2001, Hands2006-mh}, which is given by $c_s^2/c^2=(1-\mu_c^4/\mu^4)/(1+3\mu_c^4/\mu^4)$, in the BEC phase. 
According to Ref.~\cite{Son_2001}, the prediction of ChPT is valid in a low $\mu$ regime where $\mu$ is smaller than $m_V$.
Furthermore, the quark mass is still heavy in our simulations. To find the reason why ChPT shows such a nice consistency with
the lattice data will be interesting future work not only in the context of EoS but also in the context of mass
spectrum~\cite{Hands:2007uc,Wilhelm:2019fvp, Murakami}.
We also find that $c_s^2/c^2$ is larger than $1/3$, which is the value in the relativistic limit, 
at higher densities
than the regime where the BEC-BCS crossover occurs.
Eventually, our data  seem to peak around $\mu \approx m_{PS}$ and,  as density increases further, decrease so as to go away from the ChPT prediction.
{\it Such a peak of the sound velocity is a characteristic feature previously unknown from any lattice calculations for QCD-like theories.}
For example, in the finite temperature case, the sound velocity monotonically increases in $T>T_c$ and approaches 
 the relativistic limit as the temperature increases~\cite{Borsanyi:2013bia,HotQCD:2014kol}.

It is strongly believed that at ultrahigh density, $c_s^2/c^2$ approaches the relativistic limit.
Then, there arises a question of
{\it how} it approaches $1/3$.  
According to the pQCD analysis (see Appendix~A in \cite{Kojo2021-on}), it scales as $c_s^2/c^2 \approx (1- 5\beta_0 \alpha_s^2/(48\pi^2))/3$, where $\beta_0 = (11N_c -2 N_f)/3$ denotes the $1$-loop coefficient of the beta-function.
Thus, $c_s^2/c^2$ approaches the asymptotic value from {\it below}. 
On the other hand, a result based on the resummed perturbation theory suggests that $c_s^2/c^2$ approaches 
the limit from {\it above}~\cite{Fujimoto2020-bh}. 
In the numerical simulations, the maximum value of $\mu$ is limited by $\mu \ll 1/a$ to avoid the strong lattice artefact. Otherwise, the hopping term of fermions would be partially suppressed by the factor $e^{-a\mu}$  in the Wilson-Dirac operator. For the extension to larger chemical potential, we need to perform the smaller lattice spacing or lighter quark mass simulations. 
Furthermore, to obtain $c_s$ at $T=0$,
it is also required to see the EoS in the lower temperature regime by carrying out the larger volume simulations.

According to Ref.~\cite{Kojo2021-mg}, a peak of $c_s^2$ appears due to the development of the quark Fermi sea just after the saturation  of low momentum quarks.
The density at which the peak appears in our results is apparently low, i.e., $\mu \approx m_{PS}$,
but seems sufficiently high that the quark Fermi sea would be fully developed.
It supports the predictions from several effective models  based on the presence of the quark Fermi sea~\cite{McLerran2019-qh, Kojo2021-mg, Kojo2021-wh}. 
Furthermore, it is reported that the peak of sound velocity emerges around BEC-BCS crossover also in condensed matter systems  with finite-range interactions~\cite{Tajima:2022zhu}.
To ask whether or not the emergence of the peak structure is a universal property of superfluids in a BEC-BCS crossover regime, it would be important to investigate the origin of this structure as 
another future work. 
If the peak of sound velocity would be a universal property even for real $3$-color QCD as discussed in Refs.~\cite{Kojo2021-mg, Kojo2021-wh}, then  it will change one of the conventional pictures that explain the presence of massive neutron stars, namely, a first order transition
from stiffened hadronic matter to soft quark matter.

\begin{acknowledgments}
We would like to thank T.~Hatsuda,  T.~Kojo, T.~Saito, D.~Suenaga, H.~Tajima and H.~Togashi for useful conversations.
We are greatful to S.~Hands and J.-I.~Skullerud for calling our attention to erroneous data in the earlier version of the manuscript.
The consistency with ChPT was kindly suggested by N.~Yamamoto.
E.~I. especially thanks T.~Kojo T.~Hatsuda and H.~Togashi for fruitful discussions about the origin of peak, the pQCD analysis and the correspondence between the lattice data and neutron-matter analysis.
Discussions in the working group ``Gravitational Wave and Equation of State" in iTHEMS, RIKEN was useful for completing this work.
The work of E.~I. is supported by JSPS KAKENHI with Grant Number 19K03875, JST PRESTO Grant Number JPMJPR2113 and JSPS Grant-in-Aid for Transformative Research Areas (A) JP21H05190,  and the work of K.~I. is supported by JSPS KAKENHI with Grant Numbers 18H05406 and 18H01211.
The numerical simulation is supported by the HPCI-JHPCN System Research Project (Project ID: jh220021).

\end{acknowledgments}

\bibliographystyle{utphys}
\bibliography{2color}

\end{document}